**BioChemInsight: An Online Platform for Automated Extraction of Chemical Structures and Activity Data from Patents**

Zhe Wang†[1,2], Fangtian Fu†[3], Wei Zhang†[1,3], Lige Yan[4], Nan Li[4], Wenxia Deng[4], Yan Meng[1], Jianping Wu[1,5], Hui Wu[2], Wenting Wu[6], Gang Xu*[1], Xiang Li*[7], Si Chen*[4]

[1]Institute of Bioengineering, College of Chemical and Biological Engineering, Zhejiang University, Hangzhou, China

[2]Huadong Medicine Co., Ltd., 866 Moganshan Road, Hangzhou 310011, China

[3]Hangzhou VicrobX Biotech Co., Ltd., Hangzhou, China

[4]School of Medicine, Shanghai University, Shanghai, China

[5]ZJU-Hangzhou Global Scientific and Technological Innovation Center, Hangzhou, China

[6]Key Laboratory of Modern Chinese Medicine Preparations Ministry of Education, Jiangxi University of Chinese Medicine, Nanchang, Jiangxi, China

[7]School of Pharmacy, Second Military Medical University, Shanghai 200433, China

*Please address correspondence to Dr. Si Chen at caroline-sisi-chen@hotmail.com, Dr. Xiang Li at xiangli@smmu.edu.cn and Dr. Gang Xu at xugang_1030@zju.edu.cn

**Abstract**

The automated extraction of chemical structures and their corresponding bioactivity data is essential for accelerating drug discovery and enabling data-driven research. Current optical chemical structure recognition tools lack the capability to autonomously link molecular structures with their bioactivity profiles, posing a significant bottleneck in structure-activity relationship analysis. To address this, we present BioChemInsight, an open-source pipeline that integrates DECIMER Segmentation with MolNexTR for chemical structure recognition, GLM-4.5V for compound identifier association, and PaddleOCR combined with GLM-4.6 for bioactivity extraction and unit normalization. We evaluated BioChemInsight on 181 patents covering 15 therapeutic targets. The system achieved an average extraction accuracy of above 90% across three key tasks: chemical structure recognition, bioactivity data extraction, and compound identifier association. Our analysis indicates that the chemical space covered by patents is largely complementary to that contained in established public database ChEMBL. Consequently, by enabling systematic patent mining, BioChemInsight provides access to chemical information underrepresented in ChEMBL. This capability expands the landscape of explorable compound-target interactions, enriches the data foundation for quantitative structure-activity relationship modeling and targeted screening, and reduces data preprocessing time from weeks to hours. BioChemInsight is available at https://github.com/dahuilangda/BioChemInsight.

## Graphical Abstract

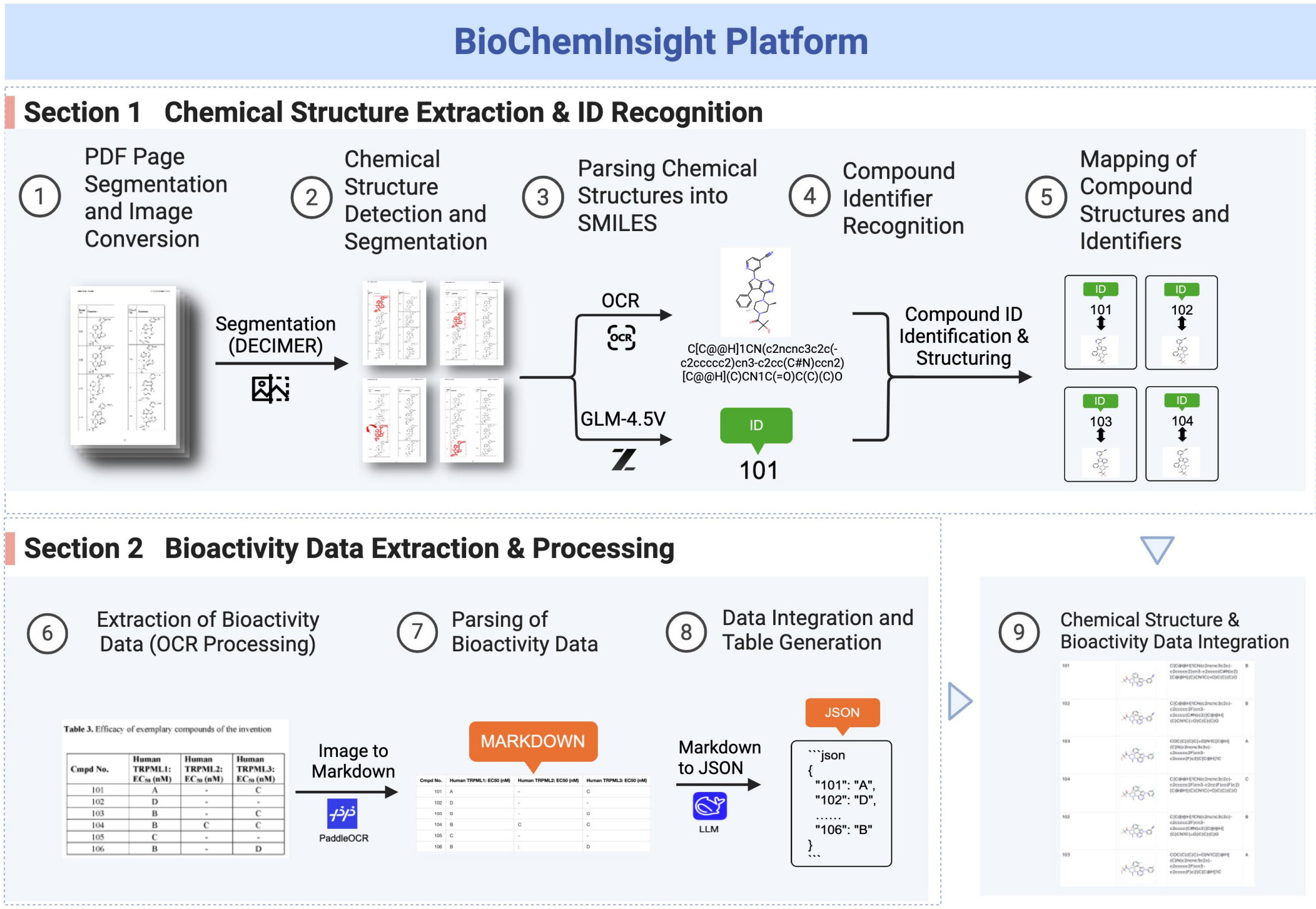

## Introduction

Recent advances in literature mining technologies have accelerated discoveries in chemistry[1–3]. Systematically extracting chemical data from literature and standardizing it into machine-readable formats serves not only chemistry but also interdisciplinary research in biomedicine, materials engineering, and systems biology, regardless of manual or automated approaches[4,5]. However, manual extraction remains time-consuming, error-prone, and labor-intensive.

Optical Chemical Structure Recognition (OCSR) automates the conversion of chemical depictions in scientific documents into machine-readable formats, overcoming the limitations of manual extraction through superior processing speed, objective interpretation, and scalable document analysis[6,7]. However, the field of OCSR faces three fundamental challenges: (1) the inability of traditional rule-based systems to accurately process diverse chemical depictions, (2) growing volumes of unprocessed scientific literature, and (3) the need for practical applications in research workflows.

Deep learning has addressed these limitations through specialized solutions: DECIMER.ai overcomes publication variability with its proprietary segmentation-classification pipeline[3], while ChemReco solves hand-drawn diagram recognition with 96.9% accuracy via EfficientNet-Transformer hybridization[8]. MolScribe tackles stereochemical complexity through geometric analysis[9], and MolMiner's integrated framework (MobileNetv2/YOLOv5/EasyOCR) delivers benchmark-leading performance with batch PDF processing - directly addressing literature backlogs[10]. Supplementary advances include ChemGrapher's bond-order precision[11], SwinOCSR's 98.58% recognition rate[12], and ABC-Net's efficient molecular reconstruction via atomic center-point detection[13]. This generation of tools collectively enables: (i) accurate digitization of legacy publications, (ii) efficient processing of contemporary literature, and (iii) seamless integration with research pipelines through real-time editing interfaces.

Despite these remarkable technological achievements, current OCSR systems exhibit a critical functional limitation: the inability to autonomously associate recognized molecular structures with their corresponding bioactivity profiles. This capability gap fundamentally constrains applications in drug discovery, where the integration of structural and pharmacological data is essential for target identification, rational compound design, high-throughput drug screening, and mechanism elucidation. The development of next-generation OCSR platforms capable of bridging this structure-activity divide represents both a significant challenge and opportunity for the field.

To operationalize this capability, we present BioChemInsight—an integrated pipeline that combines chemical structure recognition with bioactivity data extraction to automatically extract and associate structural information with bioactivity data from publications, enabling the subsequent derivation of structure-activity relationships. Our system employs a multi-component pipeline: (1) DECIMER Segmentation for accurate chemical structure detection from PDFs[3], followed by conversion to SMILES format using MolNexTR, (2) A Vision-Language Model (GLM-4.5V) processes red-boxed regions annotated by DECIMER Segmentation, containing structures and their labels, to associate each structure with its identifier through spatial correlation, and (3) PaddleOCR v2.6 converts PDFs to Markdown, after which GLM-4.6 parses this Markdown to extract specified bioactivity data (e.g., $IC_{50}$, $EC_{50}$, Ki) with normalized units (nM/μM). BioChemInsight generates structured data files that systematically associate compound IDs with their corresponding SMILES structures and bioactivity metrics, providing ready-to-use datasets for drug optimization, drug screening, and machine learning applications—all without requiring human intervention.

The complete BioChemInsight suite is publicly hosted on GitHub (https://github.com/dahuilangda/BioChemInsight), enabling easy deployment and modular customization. By automating the association between chemical structures and bioactivity profiles, the platform reduces data preprocessing time while generating high-quality datasets. These curated datasets provide high-quality training data that improves the predictive accuracy of machine learning/deep learning models and enables robust structure-activity relationship analysis. BioChemInsight represents a significant step toward data-centric pharmaceutical research by enabling end-to-end automation of structure-activity data extraction.

## Methods

### 1. BioChemInsight framework overview

BioChemInsight is a computational suite built with a React frontend and a Python 3.10 backend, designed for the automated extraction of structure-activity relationship (SAR) data from chemical literature. The system implements a modular pipeline that integrates computer vision, natural language processing, and cheminformatics tools to transform PDF documents into structured chemical-bioactivity networks. All components are version-controlled (PyTorch 2.0.0, CUDA 11.8) with dependencies managed through Conda environments. Complete validation datasets and installation guidelines are available through the project's GitHub repository at https://github.com/dahuilangda/BioChemInsight.

## 2. Automated processing pipeline

The workflow executes through four sequential stages, each handling specific aspects of chemical information extraction.

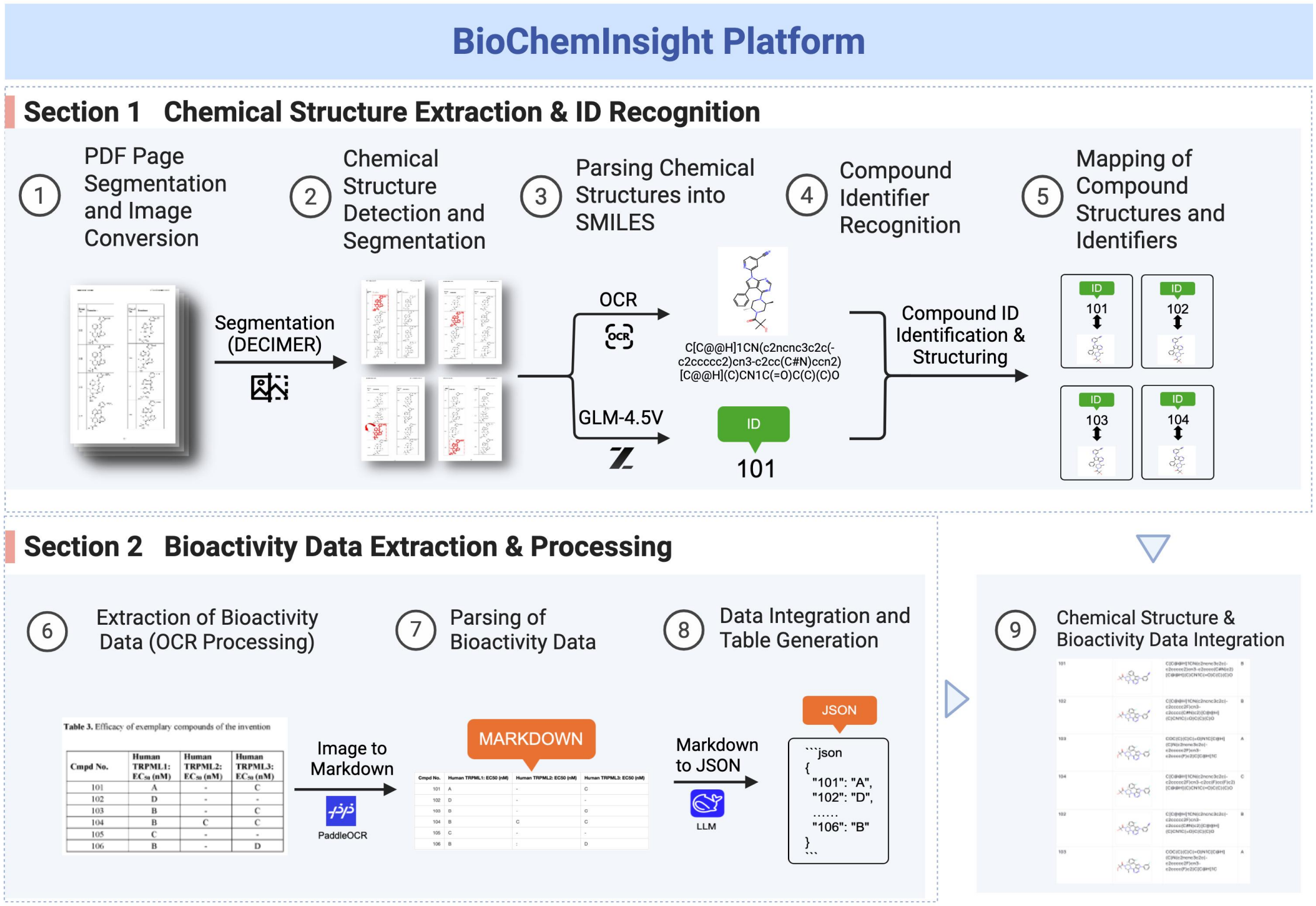

**Figure 1. Schematic overview of the BioChemInsight platform for automated extraction of structure-activity relationship (SAR) information from patent documents.** The platform consists of two core processing sections. Section 1 outlines the workflow for chemical structure extraction and compound identifier recognition, culminating in the mapping of structures to their corresponding identifiers. Section 2 depicts the pipeline for bioactivity data extraction and processing, including OCR-based text conversion, parsing via the GLM-4.5V vision-language model, and integration into structured tables. The final output combines chemical structures with associated bioactivity data to generate a comprehensive SAR dataset.

### 2.1 Document preprocessing

Input PDFs are converted into 300 DPI PNG images per page using PyMuPDF (v1.24.10), ensuring resolution adequacy for structure detection. This step normalizes document formats and handles potential PDF rendering inconsistencies across operating systems.

### 2.2 Chemical structure recognition

Chemical structure depictions are identified through DECIMER Segmentation's Mask R-CNN, which detects and crops them with 99% bounding box accuracy, while correctly excluding nearly 100% of non-chemical images[3]. The cropped structures are converted into SMILES strings by MolNexTR[14].

### 2.3 Compound identifier recognition

GLM-4.5V extracts raw compound labels (e.g., "Ex.1") from DECIMER-annotated red regions through spatial correlation. GLM-4.6 then standardizes these identifiers into consistent nomenclature (e.g., "Compound 1") using regular expression matching and synonym resolution. Finally, the system cross-validates identifiers against page coordinates and filename metadata to ensure precise matching between standardized IDs and their corresponding SMILES strings.

### 2.4 Bioactivity data extraction

PaddleOCR v2.6 converts PyMuPDF-derived PNGs into structured Markdown. Gemini-2.0-flash is then applied to extract bioactivity metrics ($IC_{50}$/$EC_{50}$/Ki) from both tabular data and contextual mentions (e.g., '$IC_{50}$ = 12.5 μM for Compound 5'). The system implements pattern matching for value extraction and contextual analysis for unit normalization (nM/μM resolution), ensuring standardized output of both textual content and quantitative bioactivity data.

### 2.5 Data integration and output

The data integration phase directly matches SMILES strings with standardized compound identifiers and pairs them with raw bioactivity values to generate a structured matrix without post-processing. The final output combines SMILES, normalized IDs, bioactivity measurements ($IC_{50}$/$EC_{50}$/Ki), and source metadata (page numbers/images), exported in CSV/JSON formats for batch processing across document collections.

## 3. Dataset Construction for Cross-Source Validation

We compiled a specialized dataset comprising 181 patents associated with 15 therapeutic targets, selected from the Cortellis Drug Discovery Intelligence database based on inclusion criteria (e.g., ⩾20 chemical depictions, small-molecule claims). Using BioChemInsight, we systematically extracted compound identifiers, SMILES strings, and biological activity values (including $IC_{50}$, $EC_{50}$, and Ki). All extracted activity values were then normalized to the nanomolar (nM) scale and underwent 100% manual validation by two independent medicinal chemists to ensure data accuracy.

### 4. Statistical analysis and output

The final matrix integrates SMILES, standardized IDs, and bioactivity values ($IC_{50}$/$EC_{50}$/Ki), with fields for page numbers and source images. Outputs are exported as CSV/JSON for downstream analysis, supporting batch processing across document repositories. Data preprocessing and statistical analysis were implemented in Python 3.10 using pandas (v1.4.2). Results were visualized via matplotlib (v3.10.0) and seaborn (v0.13.2).

## Results and Discussion

### 1. Front-end Interface and Workflow for Interactive SAR Data Extraction

The BioChemInsight Workbench's front-end interface empowers researchers to curate SAR data from patents through a modular, interactive workflow that emphasizes manual, page-level control. As shown in Figure 2A, a four-step navigation (Upload PDF → Structures → Bioactivity → Review) guides users with progress tracking (e.g., upload percentage) and clear step labels, while automatic thumbnail generation for the uploaded PDF reduces technical overhead in preprocessing. Figure 2B and 2C highlight the platform's focus on user-driven page curation. In the "Structure Explorer" (Figure 2B), users manually select pages that contain chemical structures, which are presented as grid-based thumbnails with clear page numbering. In the "Bioactivity Page Selection" module (Figure 2C), users identify pages holding bioactivity data (e.g., assay tables) and manually enter corresponding assay names. This design acknowledges the heterogeneous layout of patent documents, where chemical structures, text, and tables are often scattered; manual page-level selection ensures that relevant content is accurately targeted—a critical prerequisite for high-quality SAR extraction.

The visual layout of both modules (grid-based thumbnails, clear labeling) enables efficient browsing of multi-modal patent content. By separating structure-page and bioactivity-page selection into dedicated modules, the interface reduces cognitive load and supports precise, step-wise curation. For instance, a researcher working on a kinase-inhibitor patent can use the Structure Explorer to isolate pages showing compound structures, then switch to the Bioactivity Page Selection to locate pages containing dose-response or enzyme-activity data, and finally input assay names (e.g., “Kinase Inhibition Assay”) to contextualize the extracted values. While this manual curation approach gives users full control, it may introduce inter-user variability or become tedious for large document sets. Future versions could incorporate AI-assisted page classification (e.g., automatically detecting structure-rich or assay-rich pages) to complement—rather than replace—user judgment. Overall, the interface’s explicit manual page selection (as embodied in Figure 2B and 2C) ensures that SAR curation remains both user-directed and visually intuitive, laying a robust foundation for subsequent structure extraction and bioactivity validation.

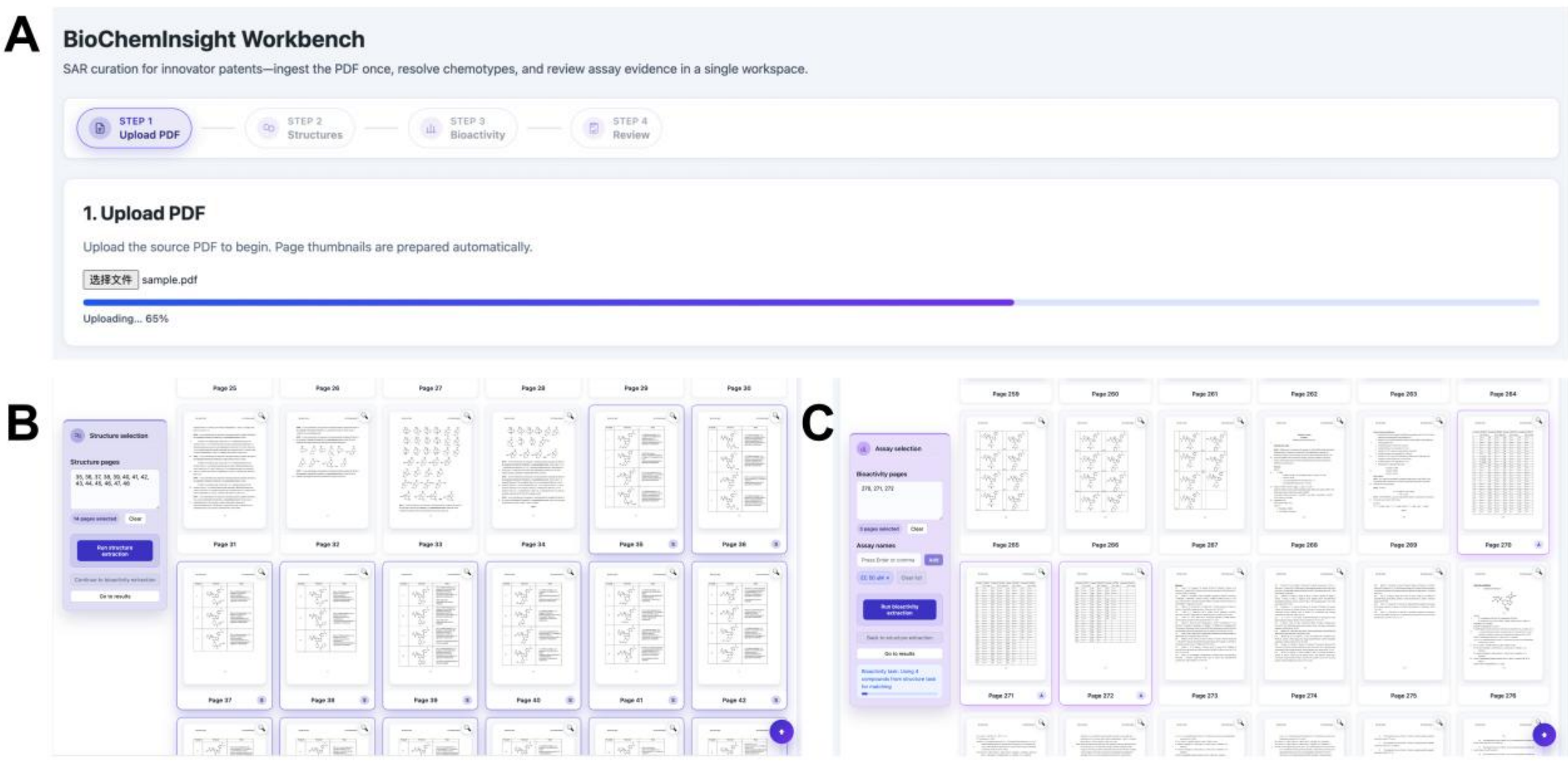

**Figure 2 User Interface of the BioChemInsight Workbench for Interactive Curation of Structure-Activity Relationship (SAR) Data from Patent Documents.** The interface streamlines the workflow from PDF ingestion to data review via modular navigation and visual components. (A) displays the four-stage workflow (Upload PDF → Structures → Bioactivity → Review) with a progress indicator for PDF upload (e.g., “Uploading: 65%”) and automatic thumbnail generation for source pages. (B) shows the “Structure Explorer” module, where users manually select pages containing chemical structures (visualized as thumbnails). After manual page selection, the system automatically extracts the chemical structures. (C) presents the “Bioactivity Page Selection” module, where users manually identify pages with bioactivity data (e.g., assay tables) and input assay names. Following page identification, the system automatically extracts the corresponding bioactivity information. This integrated workspace enables sequential navigation and manual curation of chemical structures and bioactivity evidence at the page level,

supporting efficient SAR dataset construction.

The Review interface (Figure 3) provides the final, human-supervised checkpoint in the BioChemInsight workflow, designed to ensure the accuracy of extracted SAR data before export. As shown in figure 3A, it offers retrospective control (“Adjust extractions”) and one-click export functionality. Its core component is an interactive data table that displays Compound IDs, source and extracted structures (preserved in their original patent orientation for reliable visual cross-checking), and editable bioactivity values. This design enables efficient validation; for instance, a user can swiftly confirm that an extracted structure matches its hand-drawn counterpart in the patent PDF. As illustrated in Figure 3B, users can also manually edit any incorrect structure directly within the table to ensure data fidelity.

However, the current validation occurs primarily at the end of the workflow. A potential limitation is that errors originating from earlier steps (e.g., incorrect page selection or structure mis-extraction) are only detectable at this review stage, which may require users to navigate back to previous modules for correction. Future implementations could enhance efficiency by integrating real-time, inline validation cues. For example, by automatically highlighting discrepancies between an extracted structure and its source image directly within the review table, allowing for immediate correction. Despite this opportunity for improvement, the present interface, with its editable fields, customizable layout, and quality-control annotations, establishes a robust foundation for manual verification, effectively marrying automated extraction with expert-driven data curation.

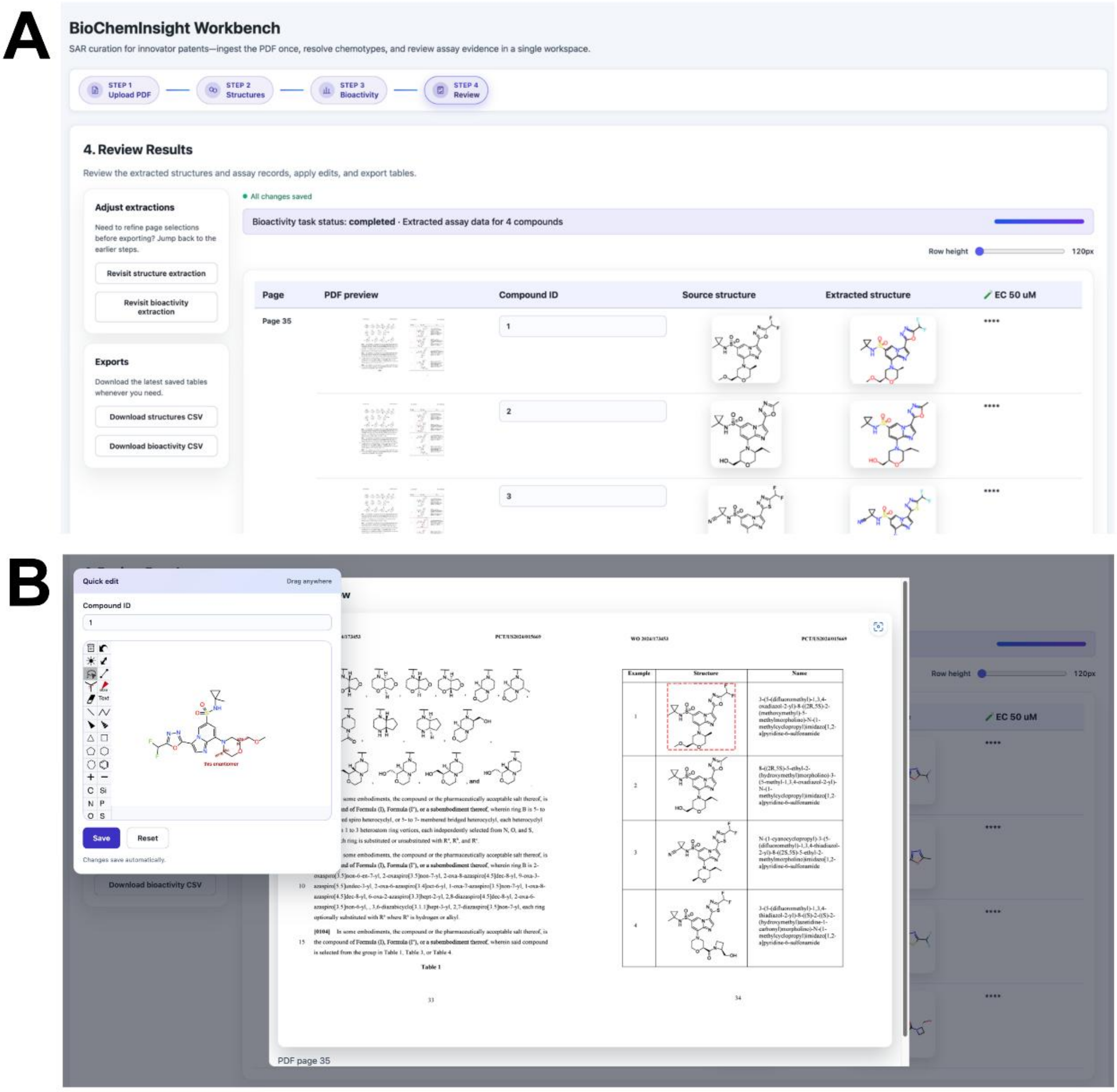

**Figure 3 Manual Validation and Editing Interface of the BioChemInsight Workbench for Curated SAR Data.** This interface enables users to inspect, refine, and export structured SAR results following initial extraction. (A) The "Review Results" panel allows users to adjust page selections via the "Adjust extractions" module (e.g., revert or reassign pages) and export curated datasets with one-click downloads (CSV/Excel). The main data table displays compound IDs, source structures, extracted structures (preserved in original patent orientation for cross-checking), bioactivity values (e.g., $EC_{50}$ in μM), and editable fields for manual correction. (B) Users can interactively edit entries for customized viewing. Together, these features support rigorous, user-driven quality control over extracted SAR data.

## 2. Performance of BioChemInsight in Extracting SAR Information Across Therapeutic Targets

As shown in Figure 4, BioChemInsight demonstrated robust and consistent performance in extracting SAR data from patents across 15 diverse therapeutic targets. Across all targets, the system achieved high extraction accuracy (>0.8) for chemical structures, biological assay data, and compound identifiers, with most metrics exceeding 0.9. Notably, several biological assays reached 100% accuracy, underscoring the tool's reliability in capturing key bioactivity data. This high performance was achieved despite substantial variation in dataset size, as the number of patents processed per target ranged from 4 to 20, further confirming the system's scalability. Moreover, extraction accuracy did not correlate with the number of patents, indicating that performance depends primarily on the inherent characteristics of the patent content (e.g., clarity of structural depictions, uniformity of assay tables) rather than on dataset volume. These results collectively validate BioChemInsight as an effective and scalable tool for accelerating SAR data curation. The system's consistently high accuracy—including perfect extraction for selected assays—ensures reliable capture of the bioactivity data essential for establishing robust structure-activity relationships.

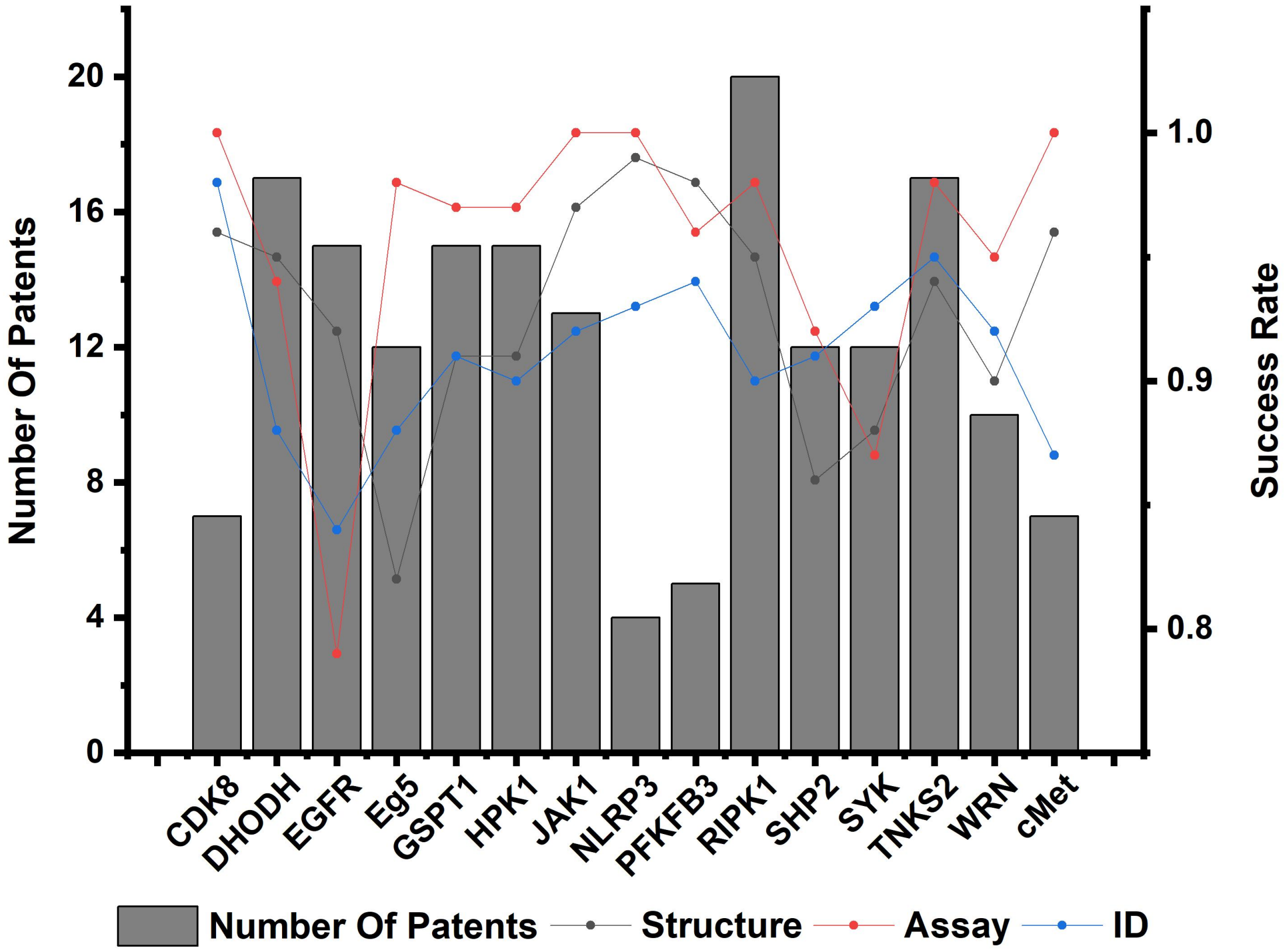

**Figure 4. BioChemInsight performance in extracting structure-activity relationship (SAR) information from patents for 15 therapeutic targets.** The bar chart (left y-axis) shows the number of patents analyzed per target. The three line plots (right y-axis) indicate the extraction accuracy of chemical structure recognition (Structure), biological assay data extraction (Assay),

and compound identifier (ID). Accuracy scores range from 0 to 1.

Figure 5 illustrates representative error cases encountered during the automated extraction of chemical information from patents using BioChemInsight. While the tool generally demonstrates competence in parsing structured data, certain systematic limitations were identified. Two primary sources of error are highlighted, as exemplified in panel (B) of the figure.First, structural extraction accuracy can be compromised by image quality issues inherent in the source documents. As seen in the provided patent excerpt (WO 2007096393), the molecular structures are presented as embedded images. In some patents, these images are subjected to blurring, down-sampling, or other forms of quality degradation, either intentionally or as a byproduct of document processing. This reduction in resolution directly impacts the performance of the optical character recognition (OCR) and structure segmentation models, leading to misinterpretations of bond connectivity, stereochemistry, or functional groups. The discrepancies highlighted by the red boxes in Figure 5B — such as misidentified functional groups or incorrect bond assignments—are often traceable to ambiguous or low-fidelity graphical representations in the original patent figures, which challenge even robust vision-based extraction algorithms.

Second, the multi-stage processing pipeline, which involves converting image-based text to structured JSON, introduces a critical point of failure in data alignment. Our methodology employs PaddleOCR for text recognition, followed by a large language model (GLM-4.6) for structuring the output into JSON. As shown in the patent table excerpt, the association between a compound identifier (ID), its molecular structure, and its corresponding assay data (Assay) is defined by their precise spatial arrangement in the original document layout. However, when OCR processing results in table cell misalignment or erroneous text block segmentation, the subsequent language model may incorrectly associate a compound's ID with another's assay value or structure. This parsing error propagates through the pipeline, generating inaccurate or nonsensical JSON records where the semantic relationships between data fields are lost or corrupted, ultimately affecting the reliability of the extracted bioactivity dataset.

These findings underscore a fundamental challenge in chemical patent mining: the reliance on converting unstructured or semi-structured document layouts into fully structured, machine-readable data. Future improvements to BioChemInsight will need to address these specific vulnerabilities. This could involve developing pre-processing algorithms to enhance low-quality chemical images, implementing more sophisticated layout analysis models that are specifically trained on diverse patent table formats, and designing post-extraction validation heuristics to cross-check the consistency between extracted structures, identifiers, and numerical assay values. Addressing these limitations is crucial for achieving high-fidelity, fully automated extraction suitable for large-scale chemical information retrieval and analysis.

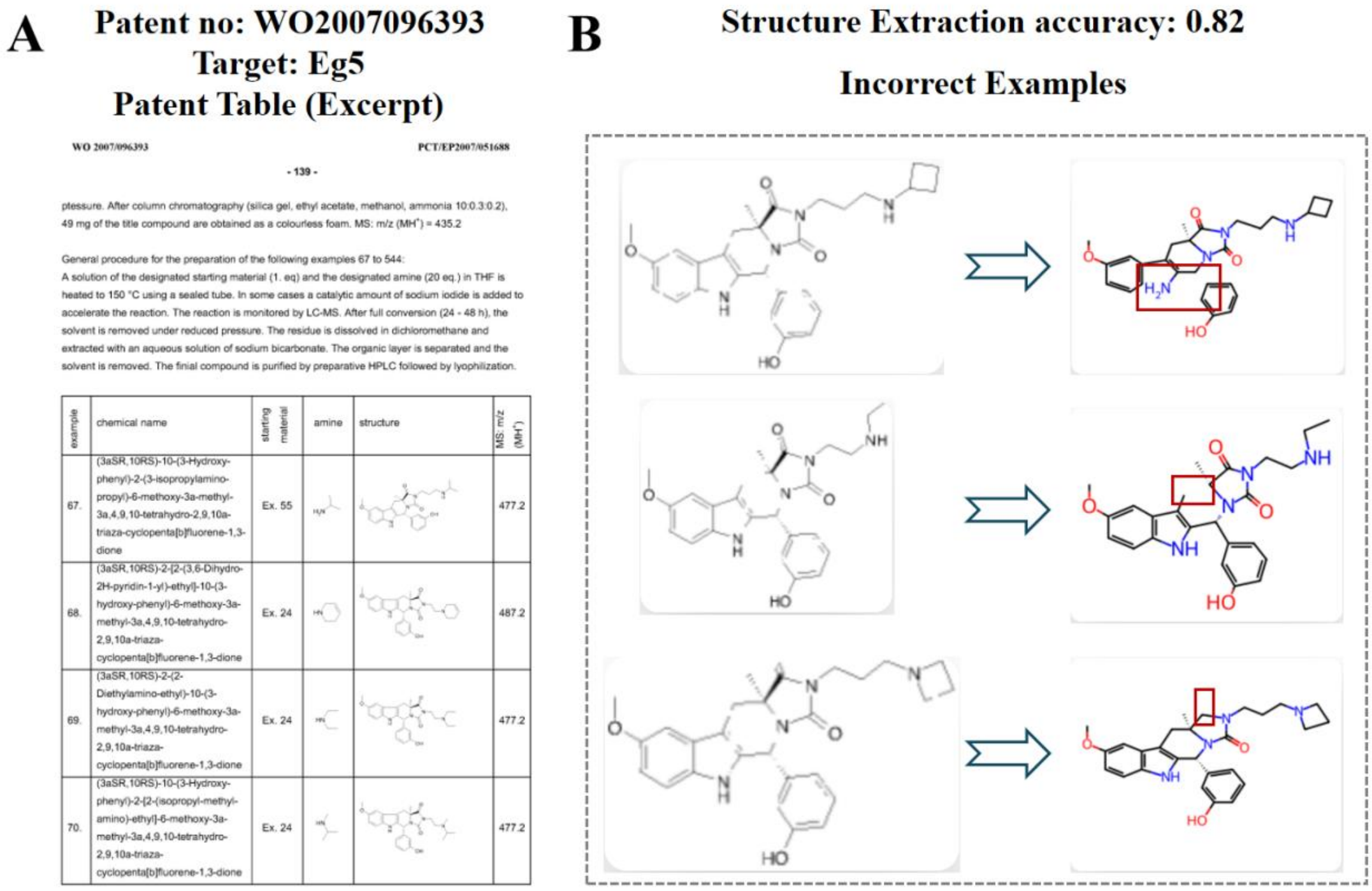

**Figure 5. Representative error cases in chemical structure extraction from patents by BioChemInsight.** (A) A representative excerpt from patent WO 2007096393, showing a table that lists compound identifiers, molecular structures, and assay activity data for Eg5 inhibitors. (B) Three representative failure cases in structure extraction. For each compound, the Original Structure (left) from the patent is compared with the Extracted Structure (right) generated by BioChemInsight. Discrepancies are highlighted with red boxes, illustrating typical error modes in automated parsing, such as misidentification of functional groups or bond connectivity.

## 3. Patent-Derived Chemical Space Exhibits Strong Complementarity to ChEMBL database

To assess the novelty and complementarity of patent-derived structure-activity relationship (SAR) data, we compared compound-target associations extracted by BioChemInsight with those from ChEMBL[15], a major public repository of bioactive molecules and SAR data. For each target, compounds from both sources were pooled, encoded as Morgan fingerprints (radius = 2,2048 bits), and visualized via dimensionality reduction using Uniform Manifold Approximation and Projection (UMAP). As shown in Figure 6, patent-derived compounds (red dots) and ChEMBL-derived compounds (blue dots) formed distinct, minimally overlapping clusters in chemical space—a trend consistent across all 15 targets analyzed. This indicates that the chemical space associated with a given target in the patent literature is substantially different from that represented in public databases.

This divergence stems from fundamental differences in data sources and curation mechanisms. ChEMBL, as a manually curated, high-quality, and large-scale open database of bioactive molecules, primarily aggregates data from peer-reviewed publications and publicly disclosed bioactivity assays. The patent-derived compounds included in ChEMBL are typically representative subsets rather than exhaustive extractions, constrained by its manual curation workflow. In contrast, patent literature itself serves as a primary source of novel chemical matter, often containing proprietary compounds that are structurally distinct and not yet deposited in public repositories. This is especially evident for emerging targets (e.g., NLRP3, PFKFB3), where patent disclosures commonly precede comprehensive academic reporting by several years.

Thus, patents and ChEMBL cover complementary regions of chemical space. This underscores the value of systematic patent mining in early-stage drug discovery: BioChemInsight facilitates access to chemical information underrepresented in established public databases. By doing so, it expands the landscape of explorable compound-target interactions and enriches the data foundation for subsequent quantitative structure-activity relationship (QSAR) modeling and targeted screening efforts.

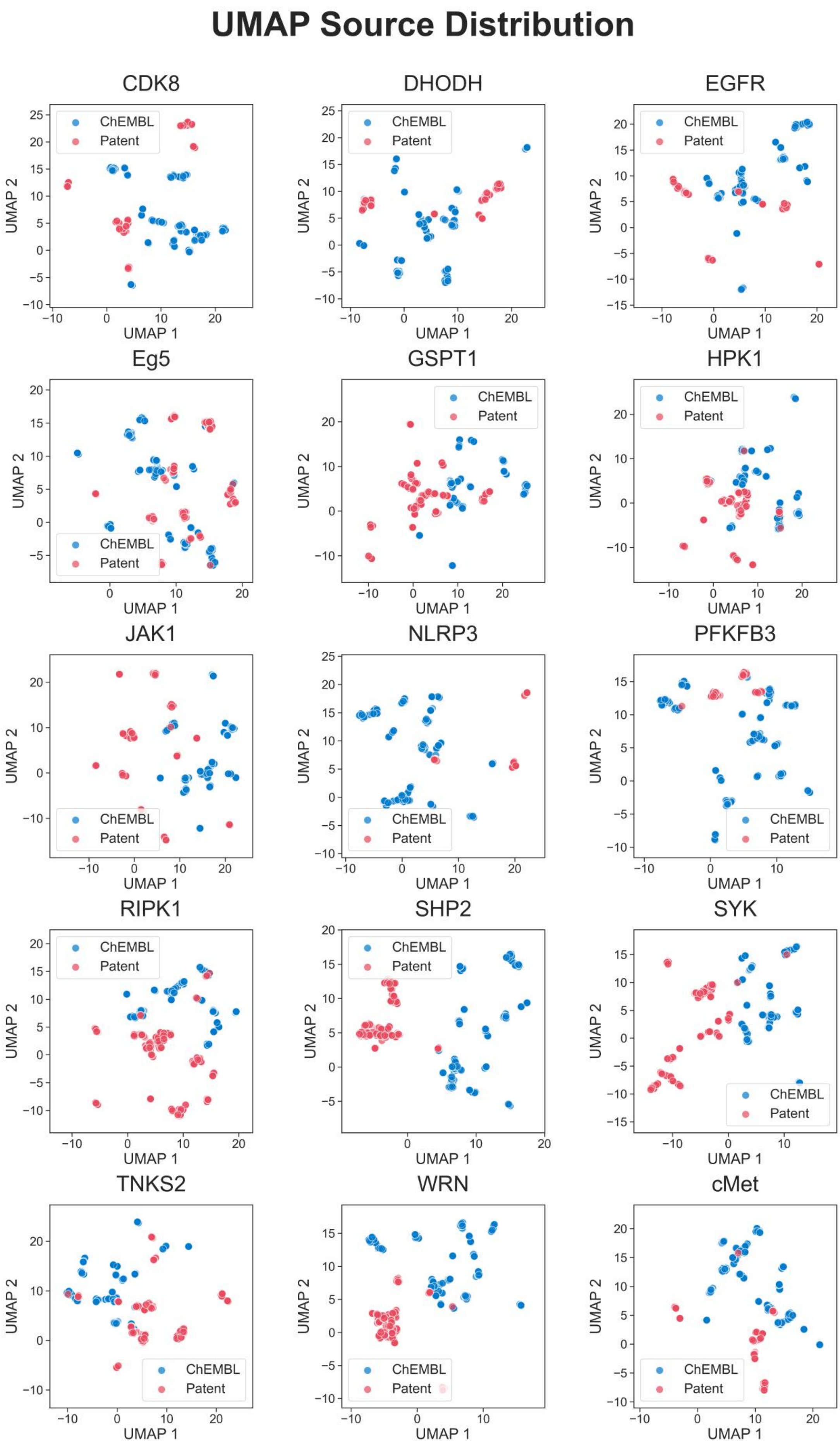
UMAP Source Distribution
CDK8
DHODH
EGFR
Eg5
GSPT1
HPK1
JAK1
NLRP3
PFKFB3
RIPK1
SHP2
SYK
TNKS2
WRN
cMet
ChEMBL
Patent
UMAP 1
UMAP 2

**Figure 6. Comparison of target representation spaces derived from BioChemInsight patent data and the ChEMBL database.** Each point corresponds to a therapeutic target, projected into two dimensions using UMAP (UMAP1 and UMAP2) based on feature vectors generated from Morgan fingerprints of compounds associated with each target. Red circles represent targets characterized from patent documents processed by BioChemInsight. Blue circles represent targets characterized from curated bioactivity records in the ChEMBL database. The minimal overlap between the two clusters indicates substantial differences in the chemical and biological information captured by patent literature versus the ChEMBL database.

## Conclusion

BioChemInsight represents a significant advancement in automated literature mining for early-stage drug discovery. It provides a unified, open-source pipeline that integrates state-of-the-art models for optical chemical structure recognition, compound identifier association, and bioactivity extraction with unit normalization, effectively converting unstructured patent documents into machine-readable SAR datasets. Our systematic evaluation on 181 patents across 15 therapeutic targets demonstrates the pipeline's robust performance, achieving an average extraction accuracy of 90% for structure, bioactivity, and identifier recognition. Crucially, analysis indicates that the chemical information extracted from patents occupies a space largely complementary to that of established public repositorie ChEMBL. Consequently, BioChemInsight unlocks access to previously underrepresented chemical and bioactivity data, thereby expanding the landscape of explorable compound-target interactions and enriching the foundation for quantitative modeling and targeted screening. By reducing data preprocessing time from weeks to hours and requiring minimal user intervention, BioChemInsight is well-suited for high-throughput applications, including automated SAR dataset construction, knowledge graph expansion, and data-driven hit identification. The pipeline is publicly available at https://github.com/dahuilangda/BioChemInsightto support reproducible, scalable chemical data mining.

## Code availability

The source code for BioChemInsight implementation and data analysis are available in Github (https://github.com/dahuilangda/BioChemInsight). The Git tag is v1.0.0.

## Author contributions

Zhe Wang, Si Chen, Fangtian Fu, Wei Zhang, Gang Xu, and Xiang Li developed the concepts for the manuscript and proposed the method. Zhe Wang, Fangtian Fu, Wei Zhang, Lige Yan, Nan Li, Wenxia Deng, Yan Meng, Jianping Wu, Hui Wu, and Wenting Wu collected the data. Zhe Wang, Fangtian Fu, Wei Zhang, Gang Xu, Xiang Li, and Si Chen designed the analyses and applications and discussed the results. Zhe Wang, Fangtian Fu, Wei Zhang, and Si Chen conducted the analyses. Gang Xu and Xiang Li helped interpret the results of the real data analyses. Zhe Wang, Fangtian Fu, Wei Zhang, Gang Xu, Xiang Li, and Si Chen prepared the manuscript and contributed to editing the paper.

## Funding

This study was supported by the National Natural Science Foundation of China (NSFC) under Grant No. 82404509.

## Conflict of interest statement

None declared.